Introducing multiverse analysis to bibliometrics:

The case of team size effects on disruptive research

Leibel, Christian[1]; Lutz Bornmann[2]

[1]ORCID: 0009-0006-6893-4901

*christian.leibel.extern@gv.mpg.de (corresponding author)*

Science Policy and Strategy Department, Administrative Headquarters of the Max Planck Society, Germany

Department of Sociology, Ludwig-Maximilians-Universität München, Germany

[2]ORCID: 0000-0003-0810-7091

Science Policy and Strategy Department, Administrative Headquarters of the Max Planck Society, Germany

## Abstract

Although bibliometrics has become an essential tool in the evaluation of research performance, bibliometric analyses are sensitive to a range of methodological choices. Subtle choices in data selection, indicator construction, and modeling decisions can substantially alter results. Ensuring robustness – meaning that findings hold up under different reasonable scenarios – is therefore critical for credible research and research evaluation. To address this issue, this study introduces multiverse analysis to bibliometrics. Multiverse analysis is a statistical tool that enables analysts to transparently discuss modeling assumptions and thoroughly assess model robustness. Whereas standard robustness checks usually cover only a small subset of all plausible models, multiverse analysis includes all plausible models. The benefits of multiverse analysis are illustrated by assessing the robustness of the findings reported by Wu et al. (2019), who observed that small teams tend to produce more disruptive research than large teams. While we found robust evidence of a negative effect of team size on disruption scores, the effect size depends substantially on the model specification. Our findings underscore the importance of assessing the multiverse robustness of bibliometric results to clarify their practical implications.

# 1 Introduction

Data do not speak for themselves. Data analysis requires theory and modeling assumptions (i.e. decisions on the specification of statistical models). Consequently, empirical results are the joint outcome of the data and the analytical decisions of researchers (Young, 2018). The crucial importance of researcher decisions becomes particularly apparent when different analysts arrive at different results regarding the same research question. Such is the case in the discussion between Wu et al. (2019) and Petersen et al. (2025): In an influential *Nature* article, Wu et al. (2019) found that small research teams produce more disruptive works than large research teams. Contradictory evidence was reported by Petersen et al. (2025), who claim to have refuted the hypotheses of Wu et al. (2019) by using a superior estimation strategy. Since the statistical models of Wu et al. (2019) and Petersen et al. (2025) are based on different modeling assumptions, it is difficult to compare and interpret their divergent findings. For want of a comprehensive discussion of modeling assumptions, little can be said with certainty about the true effect of team size on research output. The findings of Wu et al. (2019) seem to have important science policy implications, yet the current state of research must leave policy makers confused: Should they promote small or large teams when trying to foster disruptive research?

The uncertainty about modeling assumptions that surfaced in the articles of Wu et al. (2019) and Petersen et al. (2025) is symptomatic of a more widespread problem in empirical research. Even though it is well documented in the methodological literature that research outcomes depend at least as much on researcher decisions as on data characteristics, this knowledge is not reflected in the way research outcomes are typically reported. Credible research must offer "convincing evidence of inferential sturdiness" (Leamer, 1985, p. 308): It should be shown that research outcomes do not hinge on arbitrary researcher decisions.

This study draws on state-of-the-art methodological literature and presents multiverse analysis as a statistical tool for the analysis of model uncertainty in bibliometrics. We explain the basic principles of multiverse analyses and argue that they function as extensions of conventional robustness checks. In the empirical part of our study, we illustrate the benefits of multiverse analysis by using it to investigate the model uncertainty that became apparent in the discussion between Wu et al. (2019) and Petersen et al. (2025). We start by discussing the modeling assumptions made by Wu et al. (2019) and Petersen et al. (2025) and we construct a set of equally valid model specifications. Within this model space, we aim to uncover whether team size has a robust (negative or positive) effect on disruptive research output. In a second step, we identify influential modeling decisions that affect the research outcome. In doing so, we demonstrate that combining a thorough investigation of modeling assumptions with computational tests of model robustness increases the transparency and credibility of bibliometric research.

# 2 On model uncertainty and the assessment of model robustness

## 2.1 Multiverse analysis

When analyzing data, researchers are faced with numerous decisions. Among other things, analysts need to make decisions about the operationalization of important concepts, the selection of control variables, or the functional form of the statistical model. Our study deals with a fundamental issue of empirical research: *model uncertainty*. *Model uncertainty* arises when analysts make arbitrary modeling decisions for want of strong theoretical or statistical justifications that could guide them through the process of model specification. In other words: It is unclear which statistical model produces the estimate that is closest to the truth. Since scientific hypotheses are often vague regarding the statistical models with which they should be tested, model uncertainty is pervasive in empirical research.

Even though the model uncertainty is well-documented in the methodological literature (Chatfield, 1995; Leamer, 1983; Western, 1996), is has not yet become common practice (in bibliometrics) to systematically measure and report model uncertainty. Uncertainty about the optimal specification of statistical models means that there is a set of models $M_1, \ldots, M_K$ that can all be considered as equally valid. Since each model yields a unique estimate $b_1, \ldots, b_K$, the set of equally plausible model specifications "directly implies a multiverse of statistical results" (Steegen et al., 2016, p. 702). The total uncertainty of research outcomes can be estimated by calculating the variance of estimates $b_1, \ldots, b_K$, across all model specifications $M_1, \ldots, M_K$ (Young, 2009), as given in Eq. (1).

$$V_M = \frac{1}{K} \sum_{k=1}^{K} (b_k - \bar{b})^2 \quad (1)$$

$V_M$ is the variance of $b_K$ across all models and $\sqrt{V_M}$ is the model standard deviation. Statistical methods that measure $\sqrt{V_M}$ are referred to as *multiverse analyses*. The term "multiverse" alludes to the fact that these methods consider all estimates from the model space defined by $M_1, \ldots, M_K$. Several researchers from different fields have proposed multiverse-style methods as statistical tools for the measurement of model variance (Patel et al., 2015; Simonsohn et al., 2020; Steegen et al., 2016; Young & Holsteen, 2017). Overviews of multiverse analyses and guidance on how to conduct them are provided by Cantone and Tomaselli (2024), Del Giudice and Gangestad (2021), and Götz et al. (2024).

In addition to measuring the magnitude of model variance (via $\sqrt{V_M}$) multiverse analyses can also identify the causes of model variance via *influence statistics*. Consider two simple regression models that estimate the covariance of $X$ with $Y$:

$$Y = \beta_1 X + \varepsilon$$

**(2)**

$$Y = \beta_1^* X + \beta_2 Z + \varepsilon$$

**(3)**

Eq. (3) includes the control variable $Z$ in addition to the treatment variable team size. $\beta_1$ is the estimated effect of team size on disruption scores when no control variables are considered and $\beta_1^*$ is the estimate that includes the influence of the control variable. The difference ($\Delta\beta = \beta_1^* - \beta_1$) is the expected change in the coefficient on team size if control variable $Z$ is included in the statistical model. In this study, $\Delta\beta$ is defined as the *model influence* of control variable $Z$ (Young & Holsteen, 2017). In the case of multiple control variables, the model influence of $Z$ is defined as the marginal effect of including $Z$ in the model, obtained by testing all possible combinations of control variables. Eq. (2) and Eq. (3) illustrate the model influence of a control variable, but the same logic can be applied to various other modelling decisions such as the operationalization of $Y$ or the selection of the estimation procedure. Model influence statistics transparently communicate the degree to which research outcomes hinge on specific modelling assumptions.

Multiverse analysis is guided by the same notion as conventional robustness checks: Empirical results should only be taken seriously by scientists and policy makers if justifiable changes to the research design do not fundamentally alter the conclusions. However, standard robustness checks often do not yield informative estimates of model variance or model influence because they usually cover only a small subset of all plausible models $M_1, \ldots, M_K$. The limitations of conventional robustness checks seem unnecessary and unjustified given modern computational power. Nowadays, statistics software can easily run and visualize a massive number of statistical models (Muñoz & Young, 2018). Indeed, it seems that "scholars today are faced with an 'embarrassment of riches' in computational capacity: we have a lot more computational power than what is reflected in most journal articles" (Young, 2018, p. 1). Multiverse analysis rests on the same fundamental idea as conventional robustness checks, but it takes full advantage of contemporary computational power to increase the credibility and transparency of empirical research.

### 2.2 Lost in the multiverse: The case of the disruptive impact of small teams

We argue that multiverse analysis may – if used correctly – increase the transparency and credibility of empirical research by unveiling the degree to which research outcomes (and their policy implications) hinge on specific sets of modeling assumptions. To underpin our line of argument, we apply multiverse analysis to the case of the disruptive impact of small teams. We start with a short summary of two central studies by Wu et al. (2019) and Petersen et al. (2025).

In an influential *Nature* article, Wu et al. (2019) used bibliometric metadata to investigate how the growing dominance of (large) research teams in science (Wuchty et al., 2007) shapes research output. Wu et al. (2019) were the first to apply the disruption index in scientometrics. The disruption index is a bibliometric indicator of scientific innovation that measures betweenness centrality in citation networks (more details in Section 2.3). This index was calculated for a large sample consisting of more than 24 million papers published between 1954 and 2014. Wu et al. (2019) found that the contributions of large teams and small teams to scientific progress are characterized by a universal pattern: Whereas large teams tend to concern themselves with established scientific problems, small teams seem to be more capable of identifying new directions for scientific discovery. Wu et al. (2019, p. 382) argue that their study has important science policy implications: "These results suggest the need for government, industry and non-profit funders of science and technology to investigate the critical role that small teams appear to have in expanding the frontiers of knowledge".

The study of Wu et al. (2019) set in motion a new stream of research on disruptive epistemic innovations (Leahey et al., 2023; H. Y. Li et al., 2024; Y. Lin et al., 2023; Park et al., 2023), but it also spawned several articles that discuss the challenge of identifying disruptive research with bibliometric metadata (Bentley et al., 2023; Bornmann et al., 2020b; Bornmann et al., 2024; Deng & Zeng, 2023; Holst et al., 2024; Leibel & Bornmann, 2024a; Liang et al., 2022; Ruan et al., 2021; Wu & Wu, 2019). A major limitation of bibliometric network measures like the disruption index is their susceptibility to *citation inflation*. Citation inflation "refers to the systematic increase in the number of links introduced to the scientific (or patent) citation network each year" (Petersen et al., 2024, p. 937). Citation inflation is driven by two factors: First, due to shifts in scholarly citation practices, the average number of references cited in research articles has increased substantially over time (Dai et al., 2021; Nicolaisen & Frandsen, 2021; Sánchez-Gil et al., 2018). Second, and more importantly, the amount of scientific literature published each year has experienced massive growth in the past decades (Bornmann et al., 2021). Pan et al. (2018) estimated that the average number of citations in the science citation network is growing by about 5% annually. This means that the total number of citations in the network doubles approximately every 12 years. Ignoring the growth of citation networks over time may lead to serious errors in the measurement of scientific impact (Petersen et al., 2019). Citation inflation affects the citation networks of both research articles and patents (Huang et al., 2020) and renders the cross-temporal comparisons challenging.

Using a combination of deductive and empirical analysis, Petersen et al. (2024, p. 949) provide extensive proof that the disruption index "artificially decreases over time due to citation inflation". The severe limitations of the disruption index may have far reaching implications concerning the credibility of the results of Wu et al. (2019). In a follow up article to their 2024 publication, Petersen et al. (2025) argue that the "team effect" observed by Wu et al. (2019, p. 380) is confounded by citation inflation: In the time span of 60 years covered by the sample of Wu et al. (2019), team sizes increased over time, while citation inflation led to a systematic decrease in disruption scores. The negative association of team size with disruption scores observed by Wu et al. (2019) could be the result of omitted variable bias. After controlling for citation inflation, Petersen et al. (2025) found that disruptive impact incrementally increases with team size, directly contradicting the hypothesis of Wu et al. (2019).

Table 1: Differences between the model specifications used by Wu et al. (2019) and Petersen et al. (2025)

| | Wu et al. (2019) | Petersen et al. (2025) |
|---|---:|---:|
| Source of metadata | WoS | SciSciNet |
| Citation window | Up to 2014 | 5 years |
| Outliers | Included | Excluded |
| Minimum reference count | 1 | 10 |
| Statistical controls: | | |
| Publication years | Yes | Yes |
| Research fields | Yes | Yes |
| Citation counts | No | Yes |
| Reference counts | No | Yes |
| Author characteristics | Yes | No |

Note: "Outliers" refers to articles with an extremely high number of citations or cited references.

Even though the arguments of Petersen et al. (2025) contribute considerably to a richer understanding of citation dynamics, the debate about the true effect of team size on epistemic innovation is not yet settled. Table 1 highlights the differences between the methods of Wu et al. (2019) and Petersen et al. (2025). Petersen et al. (2025) claim that they arrived at different results than Wu et al. (2019) by controlling for citation inflation in the statistical analysis. However, the findings of Petersen et al. (2025) could also be the outcome of other features of their research design. As Table 1 shows, Petersen et al. (2025) did not replicate the citation window and the fixed effects model used by of Wu et al. (2019). The interpretation of the divergent results of Wu et al. (2019) and Petersen et al. (2025) is further complicated by the fact that the two studies were performed using different sources of bibliometric metadata: Web of Science (WoS, Clarivate) versus SciSciNet (Z. Lin et al., 2023).

The cases of Wu et al. (2019) and Petersen et al. (2025) illustrate the methodological challenges we wish to address in this article. First, Wu et al. (2019) and Petersen et al. (2025) arrived at divergent results because their studies rest on incompatible modeling assumptions. This indicates the need for a transparent and well justified set of model specifications. Second, the pioneering study of Wu et al. (2019) has relevant science policy implications, yet it remains an open question whether the study's findings will prove robust in a thorough assessment of model robustness. The key to answering both questions is a combination of computational model robustness and a detailed discussion of modeling assumptions.

# 3 Data and methods

## 3.1 Definition of the disruption index

Decisions regarding operationalization are among the most important choices researchers must make when analyzing data. In bibliometric studies, operationalization assumptions usually manifest in the definition of bibliometric indicators. If there is uncertainty about the optimal specification of a bibliometric indicator, the indicator may become a potent source of model uncertainty. For instance, a debate arose in bibliometrics some years ago about whether to use the average of ratios or the ratio of averages in calculating the field-normalized citation score (Opthof & Leydesdorff, 2010) – one of the most important indicators in bibliometrics. Model uncertainty is particularly evident in the extensive body of bibliometric research on interdisciplinarity: Wang and Schneider (2020, p. 239) analyzed the consistency of interdisciplinarity measures and found "surprisingly deviant results when comparing measures that supposedly should capture similar features or dimensions of the concept of interdisciplinarity". Similarly, there are several variants of the disruption index, which may or may not produce consistent results. In this section, we explain the calculation of the disruption index and why there is uncertainty about its optimal specification.

The disruption index is guided by the notion that scientific progress may occur either in a disruptive or in a consolidating manner. A disruptive paper creates new research trajectories that diminish reliance on preceding research. According to Lin et al. (2025a), disruptive research can be understood "as the replacement of older answers with newer ones to the same fundamental question – much like light bulbs replacing candles". For example, a disruptive study might introduce a novel method, making prior techniques less relevant. By contrast, consolidating contribute to cumulative scientific progress by refining, synthesizing or supporting existing ideas in the research literature. An often cited example of a consolidating study is the Nobel Prize winning paper by Davis et al. (1995), which provided compelling evidence for Bose-Einstein condensation. The paper reinforced the importance of prior research by providing empirical proof in support of established theories. Disruption and consolidation describe two distinct types of epistemic relationships between new scientific contributions and their predecessors.

The disruption index was first proposed by Funk and Owen-Smith (2017) and attempts to measure epistemic relationships via citation data. More specifically, the index is grounded in bibliographic coupling. Bibliographic coupling connects publications that cite the same references. The core idea of the disruption index is that bibliographic coupling links between a focal paper (FP) and its citing papers indicate historical continuity in the sense that future research still relies on the same sources of inspiration as the FP. Historical continuity (i.e. bibliographic coupling) can be interpreted as a sign of consolidating research whereas historical discontinuity (i.e. a lack of bibliographic coupling) can be seen as signifying disruptive research (Leydesdorff & Bornmann, 2021).

Concretely, the disruption index identifies three types of citing papers in the network of a focal paper (FP). $N_{F,t}$ are citing papers that cite only the FP (indicating discontinuity), whereas $t$ refers to the time of measurement (i.e. the citation window). $N_{B,t}$ denotes the number of citing papers that contain at least one bibliographic coupling link with the FP. $N_{R,t}$ captures papers that share at least one bibliographic coupling link with the FP, but did not cite the FP itself. The original definition of the disruption index that was proposed by Funk and Owen-Smith (2017) is the $DI_1$, which interprets all papers that cite the FP and at least one of its cited references as indicators of developing contributions. The $DI_1$ is by far the most commonly used variant of the disruption index, as illustrated by the fact that it was used in several prominent bibliometric studies published in *Nature* (Y. Lin et al., 2023; Park et al., 2023; Wu et al., 2019). The disruption index ranges from -1 (maximally consolidating) to 1 (maximally disruptive) and corresponds to the ratio in Eq. (4).

$$DI_{1,t} = \frac{N_{F,t} - N_{B,t}}{N_{F,t} + N_{B,t} + N_{R,t}} \quad \textbf{(4)}$$

We discuss the definition of the disruption index in detail to highlight the challenge of identifying disruptive and consolidating research contributions via citation data: The disruption index infers *epistemic relationships* from *citation relationships*. To make this inferential leap, the disruption index requires the strong assumption that citations always signal the flow of knowledge between research publications. This assumption corresponds to the normative citation theory (Merton, 1988). In reality, however, citations are made for a variety of different reasons: Some citations are rhetorical, some are critical, and many are perfunctory (Tahamtan & Bornmann, 2018a, 2019). The fact that not all citations imply epistemic relationships has important implications regarding the validity of the disruption index: The lower the proportion of "proper" citations in a citation network, the lower the signal-to-noise ratio of the disruption index (and similar measures).

As Leibel and Bornmann (2024b) point out in their literature review of research on the disruption index, researchers have proposed numerous modified variants of the $DI_1$, which attempt to address some of the limitations of citation data. Prominent examples of alterative index variants are the $mDI_1$ proposed by Funk and Owen-Smith (2017), which gives more weight to highly cited focal papers, and the $DI_5$ proposed by Bornmann et al. (2020a), which excludes weak bibliographic links (there are more details on the variants in Section 4.5). However, without knowing which citation relationships truly represent epistemic relationships, the signal-to-noise ratio of any variant of the disruption index remains largely unknown. In summary, the disruption index contributes to model uncertainty because its definition is to some degree arbitrary: There are several alternative definitions of the index that are – at least prima facie – equally defensible and they may or may not yield similar results.

In addition to specification uncertainty, analysts need to be aware of the susceptibility of the disruption index to data artefacts, in particular data artefacts caused by missing or faulty data on a publication's cited references. Because the disruption index is defined as a measure of betweenness centrality, it tends to assign inflated disruption scores to papers with very few cited references. In the extreme case that a publication has no cited references, the disruption index assumes its maximum value by definition. Since not citing any previous literature contradicts established scholarly citation practices, research articles that do not cite any sources of inspiration are probably exceptionally rare. Yet, bibliometric datasets may contain a substantial amount of papers with zero (or only very few) cited references because of faulty metadata (Holst et al., 2024). Such data artefacts have been shown to be more prevalent in some databases (e.g. SciSciNet) than in others (e.g. WoS) (Park et al., 2025). Avoiding data artefacts and acknowledging uncertainty about the optimal definition of bibliometric indicators are of crucial importance in research that relies on citation-based indicators.

### 3.2 Bibliometric metadata

We collected the data for this study from the Max Planck Society's inhouse version of the WoS that is based on data provided by the German "Kompetenznetzwerk Bibliometrie". The data infrastructure of the "Kompetenznetzwerk Bibliometrie" is described in detail in Schmidt et al. (2025). The data are derived from the Science Citation Index - Expanded (SCI-E), the Social Sciences Citation Index (SSCI), the Conference Proceedings Citation Index - Science (CPCI-S), the Conference Proceedings Citation Index - Social Science & Humanities (CPCI-SSH), and the Arts and Humanities Citation Index (AHCI). The bibliometric metadata covers research articles and citations between 1980 and 2023.

We assessed model robustness with several definitions of the disruption index (more details in Section 4.5). Thus, we selected a cohort of WoS papers to which we could apply citation windows of sufficient length to assess their long-term citation impact. In addition, we had to ensure that the references cited by our cohort of papers are covered by our data because metadata about cited references are required for the calculation of disruption scores. Due to the limitations of our database, we could not pick cohorts close to 1980. We decided to use the cohort of papers published between 2000 and 2005 because this cohort was published long after 1980 and enables the use of sufficiently long citation windows. Since the current snapshot of the data in our in-house database covers citations up to the end of 2023, we could apply citation windows of up to 17 years to papers published in 2005 and up to 22 years to papers published in 2000.

In our study, we took advantage of the WoS's features to minimize data artefacts. By offering bibliometric metadata about cited references that are not indexed in the WoS, the WoS data make it possible to quantify the number of "missing" (non-linked) cited references per articles. To avoid potential biases arising from data artefacts, we restricted our sample to research articles for which all their cited references are indexed as source items in the WoS. Out of the total number of 8,631,202 articles that are covered in our database between 2000 and 2005, merely 187,103 have only linked cited references. While the removal of articles with missing cited references drastically reduces our sample size, we ensure that we have complete bibliometric metadata on the citation networks of the articles we analyze. Since an unbiased estimate from a small sample is more desirable than a biased estimate from a large sample, the benefit of improved metadata quality outweighs the reduction in sample size. We excluded uncited articles because their disruption score is zero by definition.

Table 2: Descriptive statistics

| Variable | Minimum | Maximum | M | SD |
|---|---|---|---|---|
| Team size (IV) | 1 | 15 | 5.49 | 2.68 |
| $DI_1$ (DV) | -83.33 | 100 | -0.43 | 3.32 |
| $DI_2$ (DV) | -75.00 | 100 | 0.11 | 3.70 |
| $DI_3$ (DV) | -75.00 | 100 | 0.43 | 3.79 |
| $DI_4$ (DV) | -66.67 | 100 | 0.63 | 3.81 |
| $DI_5$ (DV) | -66.67 | 100 | 0.77 | 3.82 |
| Publication year | 2,000 | 2,005 | 2,002.96 | 1.65 |
| Citations | 1 | 134,360 | 64.25 | 402.91 |
| Cited references | 1 | 141 | 25.35 | 15.38 |

*Note.* n = 129,229; M = mean; SD = standard deviation. Citation counts and disruption scores were calculated using all citations up to the end of 2023. Disruption scores were multiplied by 100 to avoid small decimal places.

Table 2 shows descriptive statistics for the variables that are included in the regression models and in the multiverse analysis. Like Wu et al. (2019) and Petersen et al. (2025), we measured team size using the number of authors. To replicate the fixed effects regression models used by Wu et al. (2019), we linked our bibliometric metadata obtained from the WoS with author profiles from Scopus (Elsevier). Scopus assigns IDs to author profiles, which enables us to run regression models with author fixed effects. We opted to use Scopus author profiles, because they are of high quality as a result of extensive curation and quality assessment (Baas et al., 2020). We considered only authors with at least two research articles in our sample because this is the minimum number required to estimate author fixed effects. The analytical sample consists of 129,229 distinct research articles. In the fixed effects regression models, articles enter multiple times if they appear in the publication records of multiple authors. Thus, the sample used in the fixed effects regression models consists of 469,590 (non-distinct) articles by 140,735 authors.

Table 3: Distribution of articles published between 2000 and 2005 in our sample and the WoS.

| Field | Number of papers | Percent (sample) | Percent (WoS) |
|---|---:|---:|---:|
| Biology | 55,919 | 43 | 23 |
| Medicine | 45,704 | 35 | 25 |
| Physical sciences | 10,203 | 8 | 13 |
| Chemistry | 9,701 | 8 | 14 |
| Engineering | 3,457 | 3 | 1 |
| Multidisciplinary | 3,211 | 2 | 9 |
| Other | 1,034 | 1 | 15 |
| Total | 129,229 | 100 | 100 |

In this study, research fields were assigned to papers using WoS subject categories. Since Wu et al. (2019) provided their classification system upon request, this study replicates their approach used in the *Nature* article. We recoded 258 WoS subject categories into six major fields. Each journal was assigned a major field, and each research article was classified based on the field of the journal in which it was published. Major fields with less than 1,000 papers were subsumed under "Other". Table 3 shows the distribution of articles across research fields in our data and in the WoS. The analytical sample is not a random sample of research articles in the WoS, but a sample of articles that has been selected based on the WoS coverage of linked cited references (see above). Thus, the majority of the articles in our sample are from fields for which the WoS covers a large percentage of the referenced literature. About 59% percent of the articles in our sample are from the natural sciences (i.e. biology, chemistry, and physical sciences) and about 35% are from medicine. Articles from other fields and articles published in multidisciplinary journals amount to only about 6% of our sample. The differences between the proportions of fields in our sample compared to the WoS are a direct result of the WoS's coverage of the field-specific literature. Among all research fields, the WoS offers the best coverage for biology and medicine. Consequently, these two fields are overrepresented in our sample. Conversely, the social sciences and humanities are underrepresented because they are significantly less covered in the WoS (Aksnes & Sivertsen, 2019).

When selecting samples from the WoS there is a trade-off between generalizability across fields and metadata quality. Maximizing generalizability would come at the cost of a heightened risk of measurement bias because articles from fields with poor metadata quality would be included in the sample. Because the measure at the center of our study, the disruption index, is susceptible to data artefacts, we opted to maximize metadata quality instead of generalizability across fields. As a result of our sample selection criteria, the generalizability of our findings is mostly limited to articles from the natural sciences and medicine.

### 3.3 Regression models

The regression models formalized in Eq. (5) and Eq. (6) represent the statistical models used by Wu et al. (2019) and Petersen et al. (2025). Mimicking the statistical models used in the two original studies, we calculated disruption scores up to the year 2023 (i.e. the longest possible time span) in Eq. (5) and applied a five-year citation window to the citation counts and disruption scores in Eq. (6). The $DI_1$ score of publication $p$ is a function of $p$'s team size, publication year, and research field. Team size, publication year, and research field are coded as dummy variables. Additionally, the fixed effects regression model includes the author-specific error term $\alpha_i$. Like Wu et al. (2019), we clustered standard errors using author IDs in the fixed effects regressions.

$$DI_{1,2023}^{i,p} = \beta_0 + \beta_1\, Team\ size_p + \beta_2\, Publication\ year_p$$
$$+\beta_3\, Research\ field_p + \alpha_i + \varepsilon_{i,p}$$

**(5)**

Petersen et al. (2025) argue that the model used by of Wu et al. (2019) suffers from omitted variable bias because it fails to consider citation inflation. To mitigate this bias, Petersen et al. (2025) suggest controlling for the non-linear dependency of disruption scores on citation counts and reference counts. Notably, the models used by Petersen et al. (2025) does not include author fixed effects. Like Petersen et al. (2025) we only considered articles with 10 ≤ reference count ≥ 200 and citation count ≤ 200 when running Eq. (5). Reference and citation counts enter the regression models logarithmically to address the skewness of their distributions. We modeled non-linear effects by including (citation counts)$^2$ and (reference counts)$^2$. In the following Sections, Eq. (5) is referred to as the *W-specification* and Eq. (6) is referred to as the *P-specification*.

$$DI_{1,5}^{p} = \beta_0 + \beta_1\, Team\ size_p + \beta_2\, Publication\ year_p + \beta_3\, Research\ field_p$$
$$+\beta_4\, Refernce\ count_p + \beta_5\, (Refernce\ count_p)^2 + \beta_6\, Citation\ count_{p,5}$$
$$+\beta_7\, (Citation\ count_{p,5})^2 + \varepsilon_p$$

**(6)**

## 3.4 Multiverse analysis

The multiverse analysis was conducted using the *multivrs* module for the statistics software *Stata* provided by Young and Holsteen (2017).[1] When performing multiverse analyses, it is of crucial importance not overestimate model uncertainty by comparing valid models specifications with ones that are clearly misspecified. Thus, the computational assessment of model robustness must be preceded by a careful selection of equally defensible model specifications (Auspurg, 2025). The goal of Section 4 is to arrive at a well justified definition of the model space.

[1] In addition to *multivrs*, this study also uses the *coefplot* and *estout* modules provided by Jann (2005, 2014) and the graphic schemes provided by Bischof (2017).

# 4 Defining a set of equally plausible models

In the following sections, we turn to our main research objective: Estimating and reducing model uncertainty by critically evaluating the plausibility of the underlying modelling assumptions. Specifically, we aim to exclude assumptions that, upon closer examination, lack convincing justifications.

## 4.1 Defining the research goal

Clearly defined research goals are a prerequisite of solid research: It is impossible to determine the plausibility of modeling assumptions without a precise understanding of the association or the causal effect one is trying to estimate. The colloquial term "effect" is problematic because it may refer to various types of causal effects that require different and even incompatible modeling assumptions (Lundberg et al., 2021). In more precise terminology the "team effect" posed by Wu et al. (2019, p. 380) can be defined as the *total causal effect* of team size on the disruptive epistemic properties of research articles.

Clarifying the research goal is necessary to determine the appropriate selection of control variables. Total causal effects include causal mechanisms, which are modeled via mediators. Since mediators are part of the total effect of team size, mediators should not be controlled in regressions models to avoid overcontrol bias (Elwert & Winship, 2014). Furthermore, the limitations of the research design needs to be considered: The gold standard for the identification of causal effects are experimental designs, yet the studies of Wu et al. (2019) and Petersen et al. (2025) both rely on observational data, which are susceptible to *omitted variable bias* due to the lack of a randomly assigned treatment (Angrist & Pischke, 2009). To address omitted variable bias, all models in the multiverse include publication year and research field as covariates. These covariates do not contribute to model uncertainty as their inclusion is standard practice in bibliometrics.

### 4.2 Fixed effects regression models

In observational data, the characteristics of authors who tend to work in small teams may differ from the characteristics of those who work in large teams. It is possible that particularly creative authors self-select into small research teams. This self-selection can be addressed via author fixed effects, which remove person-specific variance that may contaminate the estimates from regression models (Brüderl & Ludwig, 2015). Thus, the inclusion of author fixed effects in the model specification is grounded in causal justifications. Models on the individual level without author fixed effects almost certainly suffer from omitted variable bias.

### 4.3 Controlling for citation inflation

Petersen et al. (2024) provide convincing evidence that the disruption index is biased by citation inflation. In their follow up publication, Petersen et al. (2025) propose that citation inflation should be addressed by controlling for citation counts and reference counts. While both citation counts and reference counts are undoubtably affected by citation inflation, it is also necessary to determine whether they are confounders or mediators.

Citation counts can be considered an important confounder. The main reason why Wu et al. (2019, p. 378) proposed the disruption index in bibliometrics is that “citation counts alone cannot capture distinct types of contribution”. The study of M. T. Li et al. (2024) lends further support to this argument: They found that highly cited publications do not tend to receive particularly high disruption scores. All in all, the literature on scientific innovation shows that a high citation impact is not sufficient to generate disruptive epistemic relationships. Since citation counts do not signal disruptive or consolidating epistemic relationships between publications, they should be controlled in regressions models to avoid biases related to citation inflation.

The causal role of reference counts in the model is less clear than the role of citation counts. On the one hand, the increasing tendency of authors to cite more previous works is part of citation inflation (Petersen et al., 2024), which means that the disruption index may be biased in favor of articles with a low reference count. Petersen et al. (2025) treat reference counts as a confounder and recommend controlling for reference counts to avoid biased statistical estimates. For similar reasons, articles with less than 5 or less than 10 cited references are commonly excluded in analyses involving disruption indices (Bornmann et al., 2020a; Bornmann & Tekles, 2019b; Deng & Zeng, 2023; Leibel et al., 2024; Petersen et al., 2025; Sheng et al., 2023). On the other hand, some researchers argue that there may be a meaningful relationship between the reference of authors and the disruptive impact of their papers. It is possible that researchers who have radical new ideas which were not rooted in the (field-specific) literature cannot cite many sources of inspiration (Park et al., 2025; Tahamtan & Bornmann, 2018b). Consider, for example, the paper by Hirsch (2005) that introduced the *h* index. Hirsch (2005) cited only 6 references, but the article is disruptive in the sense that it launched a new stream of research. Under the assumption that highly disruptive articles (by small research teams) are likely to cite fewer references than consolidating articles, reference counts must be considered a collider instead of a confounder.

The literature offers two competing hypotheses regarding the causal role of reference counts, but there is not enough empirical evidence to settle this debate. The analyst is thus faced with a dilemma: Controlling for reference counts may induce *endogenous selection bias* (Elwert & Winship, 2014) if reference counts are a collider, but not conditioning on reference counts risks omitted variable bias due to citation inflation. It is unclear which option is preferable.

We coded citation counts and reference counts like Petersen et al. (2025). Both variables enter the regression models logarithmically to address the skewness of their distributions. We modeled non-linear effects by including $(\text{citation counts})^2$ and $(\text{reference counts})^2$ in every model specification that contains citation counts and/or reference counts as covariates. Furthermore, we integrated the exclusion criteria used by Petersen et al. (2025) (at least 10 cited references) and Wu et al. (2019) (at least 1 cited reference) as well as an intermediate exclusion threshold that limits the sample to focal papers with at least 5 cited references.

### 4.4 The length of the citation window

Citation windows are a source of model uncertainty because citation networks change over time. However, the selection of the citation window is not entirely arbitrary. The importance of some scientific achievements is not recognized until several years after their publication (Ke et al., 2015). Bornmann and Tekles (2019a) showed that it may take a long time – in some cases more than 10 years – until articles reach stable disruption scores. The time dependence of disruption scores is relevant for the estimation of team size effects because the works of small research teams "experience a much longer citation delay" than the works of large teams (Wu et al., 2019, p. 381). Lin et al. (2025b) compared several citation windows and found that short citation windows underestimate the long-term disruptive potential of small research teams. We removed the 5-year citation window from the set of equally plausible models because the literature indicates that longer citation windows are preferable for the identification of major scientific achievements.

### 4.5 The definition of the disruption index

While there are numerous alternative definitions of the disruption index, we included only a limited number of them in the multiverse analysis. We chose this conservative approach because measures should only be compared in a multiverse analysis if they are equally valid measures of the same concept (Del Giudice & Gangestad, 2021). Thus, we only included variants of the disruption index in the multiverse analysis, for which there is empirical evidence supporting their validity as measures of disruptive impact. Specifically, we consider two sources of uncertainty in the definition of the disruption index: The time of measurement (i.e. the citation window) and the application of bibliographic coupling.

The operationalization of epistemic relationships with disruption scores is complicated by the fact that citation networks change over time. Depending on the citation window, one and the same publication may appear either disruptive or consolidating (Bornmann & Tekles, 2019a). It is unclear which citation window best reflects a paper's "true" disruptive properties. To estimate how much the choice of the citation window affects the research outcome, we ran the multiverse analysis with four different citation windows, which cover short term, medium term, and long-term citation impact. In every model specification that includes the citation count as a covariate the same citation window was applied to both the citation counts and the disruption score.

In addition to alternative citation windows, we also considered several definitions of bibliographic coupling. The original definition of the disruption index that was proposed by Funk and Owen-Smith (2017) implies that $N_{B,t}$ captures every citing paper that cited at least one of the FP's cited references. This means that even the weakest possible bibliographic coupling links (consisting of just one citation) are interpreted as indicating knowledge flow between publications. Bornmann et al. (2020a) pointed out that a higher threshold for the identification of bibliographic coupling may yield a better measure of knowledge flow by reducing the noise caused by rhetorical or perfunctory citations. They proposed the $DI_5$, which defines $N_{B,t}$ such that it only includes publications that cited at least five of the FP's cited references. The proposal of Bornmann et al. (2020a) is supported by several empirical studies on the comparative validity of different variants of the disruption index. In validation studies, the $DI_5$ (with $N_{B,t}^5$) performed at least as good and often better than the $DI_1$ (Bittmann et al., 2022; Bornmann et al., 2020a; Bornmann & Tekles, 2021; Deng & Zeng, 2023; Leibel et al., 2024; Wang et al., 2023). In light of this empirical evidence, we believe that the $DI_5$ and similar variants can be regarded as valid alternatives to the $DI_1$. We tested five definitions of bibliographic coupling in the multiverses analysis: The threshold for bibliographic coupling ranges from 1 citation ($DI_1$) to five citations ($DI_5$). We used the percentile ranks of the disruption scores in all analyses to get comparable effect sizes from all regression models.

### 4.6 Exclusion of outliers

The construction of the analytical sample is an important aspect of model uncertainty that enters the multiverse via exclusion criteria. Petersen et al. (2025) excluded outliers (reference count > 200 or citation count > 1000) from the analysis. The multiverse contains models with and without outliers to test the model influence of focal papers with extreme values. Ideally, the results of an inference should not be driven by a small number of papers and their exclusion should not substantially alter the research outcome.

### 4.7 A set of equally plausible models

Table 4 shows the model ingredients that represent the building blocks of a set of equally plausible regression models. By thoroughly discussing modeling assumptions, we identified model specifications that do not belong in a set of equally valid models. Specifically, we excluded models that do not contain author fixed effects, do not control for citation inflation via citation counts, and/or use a short citation window of only 5 years. We consider all 180 models in the model space as equally plausible. There are no convincing theoretical or statistical reasons to prefer any particular model specification over its 179 alternatives. The total causal effect of team size on disruptive research output is robust if and only if the model specifications in Table 4 yield estimates that support the same research outcome.

Note that neither the W-specification nor the P-specification are included in the set of equally plausible models. This is the case because both model specifications lack control variables that we identified as confounders in the previous sections. The W-specification does not control for citation inflation and the P-specification lacks author fixed effects.

Table 4: Set of equally plausible model specifications

| Dimension | | Specifications | |
|---|---|---|---|
| Operationalization | Disruption index | 1 – $DI_1$<br>2 – $DI_2$<br>3 – $DI_3$<br>4 – $DI_4$<br>5 – $DI_5$ | 5 |
| | Citation window | 1 – 10 years<br>2 – 15 years<br>3 – All citations up to 2023 | 3 |
| Covariates | Fixed set | 1 – Publication year, research field | 1 |
| | Citation counts | 1 – included | 1 |
| | Reference counts | 1 – included<br>2 – excluded | 2 |
| Exclusion criteria | Outliers | 1 – included<br>2 – excluded | 2 |
| | Minimum reference count | 1 – 1 cited reference<br>2 – 5 cited references<br>3 – 10 cited references | 3 |
| Model | Type of regression model | 1 – Author fixed effects regression | 1 |
| | | | 180 |

# 5 Results from a multiverse of equally plausible model specifications

## 5.1 Assessing model robustness

Figure 2 shows the range and the density of estimates from the multiverse of 180 equally valid model specifications.[2] We managed to replicate the substantial findings of Wu et al. (2019) and Petersen et al. (2025): The W-specification yields a large negative estimate (-1.05), whereas the P-specification produces a very small, but statistically significant, positive estimate (0.18).[3] The multiverse analysis thus confirms that the model specifications proposed by Wu et al. (2019) and Petersen et al. (2025) lead to contradictory research outcomes even when applied to the same data.

[2] For the sake of full transparency, Table 10 and Figure 4 (in the Appendix) describe the set of all possible models – not limited to the set of plausible models we selected in Section 4.

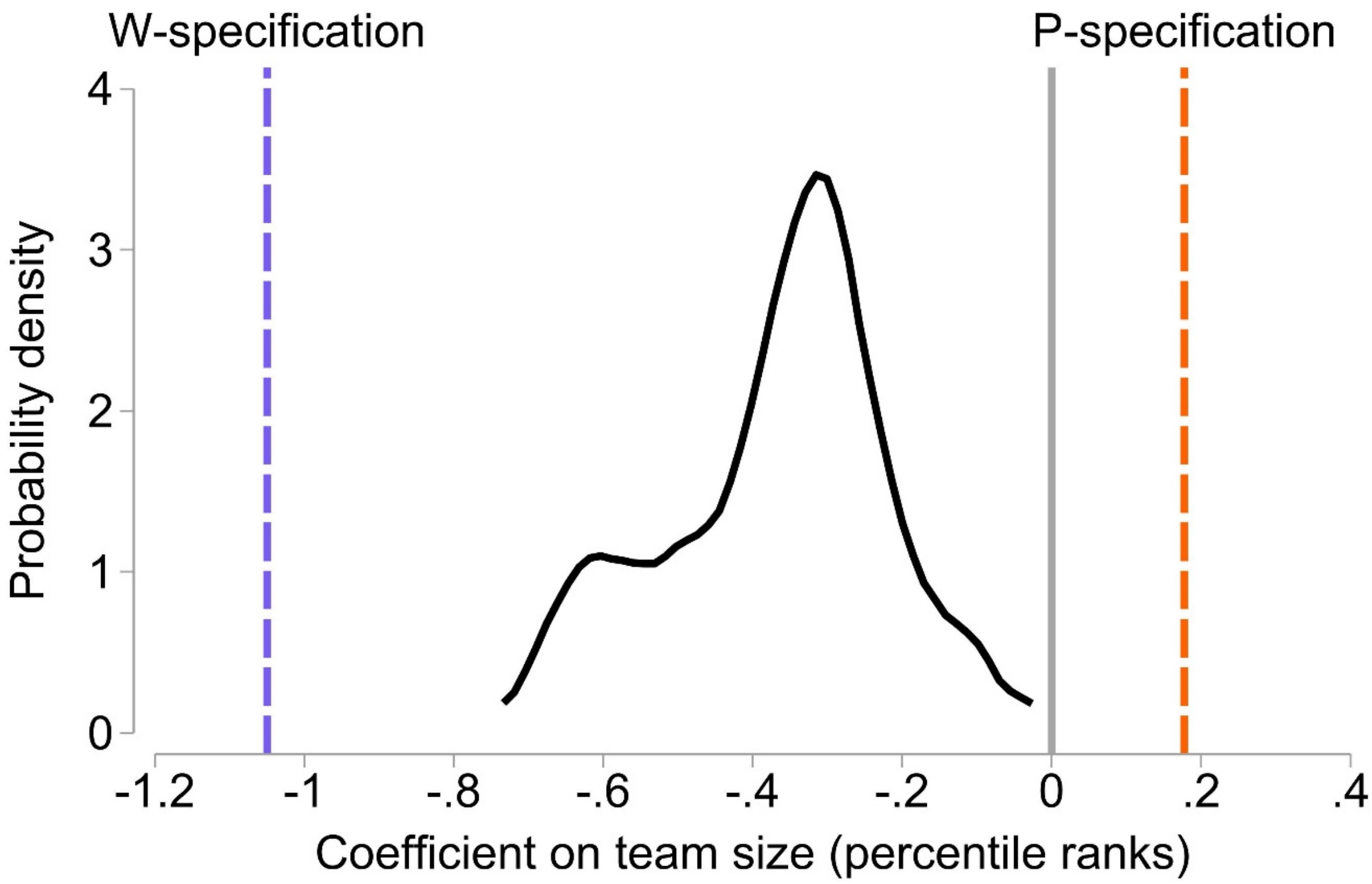

Figure 1: Probability density of the coefficients of disruption scores on team size from a multiverse of 180 equally plausible fixed effects regression models. The coefficients obtained from the W-specification and the P-specification are highlighted. Effects sizes are measured in disruption percentile ranks.

The robustness of the team effect is reflected in the range, variance, and sign stability of the estimates from the multiverse analysis. The model robustness statistics are presented in Table 5. The multiverse achieves a sign stability of 100% meaning that all estimates from the set of equally plausible models indicate a negative and statistically significant effect of team size on disruptive research output. In other words, all 180 model specifications estimate that the research of small teams has more disruptive impact than the research of large teams. Consequently, the hypothesis of Wu et al. (2019) survived our thorough assessment of model robustness.

[3] The complete regression tables for the W- and the P-specification are presented in the Appendix (Table 8 and Table 9). Figure 2 shows the coefficients from W- and the P-specification for each team size.

Table 5: Model robustness statistics

| Model space | Mean (b) | Model SD | Sign stability |
|---|---|---|---|
| 180 | -0.37 | 0.15 | 100% |

*Note*. b = coefficient on team size, SD = standard deviation. Sign stability is measured as the percentage of negative and statistically significant estimates within the model space.

Table 6: Minimum and maximum estimates from the multiverse of 180 model specifications

| DV | Citation window | Minimum reference count | Control: Reference counts | Estimate | Robust SE | p |
|---|---|---|---|---|---|---|
| $DI_5$ | End of 2023 | 1 | Excluded | -0.69 | 0.02 | < 0.001 |
| $DI_1$ | 10 years | 5 | Excluded | -0.07 | 0.02 | < 0.001 |

Note. DV = dependent variable, SE = standard error. Effect sizes are measured in percentile ranks. Standard errors are clustered using author IDs.

The variance of the estimates in the multiverse measures the stability of estimated effect sizes. The fact that the model standard deviation (0.15) corresponds to a sizeable proportion of the average estimate (-0.37) implies that the size of the team effect varies considerably across different model specifications. The range of estimated effect sizes is illustrated in Table 6, which shows the minimum and maximum estimate from the multiverse analysis.[4] The minimum estimate is only -0.07, implying that small and large research teams contribute about equally to disruptive innovation in science. The maximum estimate is about ten times larger than the minimum estimate and predicts that the disruptive impact of research articles decreases by about 6 percentile ranks when team size increases from 1 to 10 team members. An effect of this size aligns with the findings of Wu et al. (2019) and Lin et al. (2025b). In summary, we find that the multiverse covers a considerable range of effect sizes, which have different implications regarding the practical relevance of the team effect.

[4] Figure 4 in the Appendix shows the detailed dummy coefficients for the maximum, median, and minimum estimates form the multiverse.

### 5.2 Identifying remaining sources of model uncertainty

The multiverse analysis yielded a robust negative association between team size and disruptive impact, but it also revealed a considerable model standard deviation. This means that significant amount of model uncertainty remains even after restricting the model space to equally plausible models. The model influence statistics in Table 7 reveal the drivers of the remaining model uncertainty.

Our findings lend empirical support to the methodological arguments of previous studies. In line with the results of Lin et al. (2025b), we find that longer citation windows yield larger negative estimates. The disruptive potential of small teams appears to be especially pronounced when considering long-term disruptive impact. Like Petersen et al. (2025) we find that controlling for reference counts and restricting the sample to papers with a minimum number of cited references reduces the size of the team effect. The main source of model uncertainty, however, is the definition of the disruption index. The average estimate from models with the $DI_1$ is -0.19 while the average estimate from models with the $DI_5$ is -0.43. The difference between the two averages is 36% larger than the estimate from the P-specification, indicating substantial model influence. In general, index variants that only consider strong bibliographic coupling links tend to yield much larger negative estimates compared to the $DI_1$.

By providing empirical measures of model influence, the multiverse analysis yielded key insights into the consequences of modelling assumptions. While the discussion between Wu et al. (2019) and Petersen et al. (2025) mainly revolved around citation inflation, the influence statistics unveiled another highly important source of model variance: The coefficient of team size depends substantially on the specification of the disruption index. Even though we created a conservative measure of model influence by only testing five (out of numerous) index variants, the disruption index nonetheless generates a substantial amount of model variance. This finding suggests that definition of the disruption index can substantially alter research outcomes (and their policy implications), especially when dealing with small effects.

Table 7: Influence of modelling decisions on the coefficient on team size

| Benchmark: The coefficient from the P-specification is 0.1776 | | |
|---|---|---|
| Variable | Marginal effect of (variable) inclusion | Percent change from benchmark |
| Reference counts | 0.1376 | 77% |
| Exclusion of outliers | 0.0022 | 1% |
| Index variants | | |
| $DI_1$ | Reference | |
| $DI_2$ | -0.1547 | -87% |
| $DI_3$ | -0.2353 | -132% |
| $DI_4$ | -0.2578 | -145% |
| $DI_5$ | -0.2411 | -136% |
| Citation windows | | |
| 10 years | Reference | |
| 15 years | -0.0518 | -29% |
| Up to 2023 | -0.0915 | -52% |
| Minimum reference count | | |
| 1 | Reference | |
| 5 | 0.0202 | 11% |
| 10 | 0.0533 | 30% |

*Note.* Influence statistics were calculated based on 180 regression models.

# 6 Discussion

Like any type of data analysis, bibliometric analyses come with uncertainty due to "researcher degrees of freedom" (Gelman & Loken, 2014, p. 460). Ensuring the robustness of empirical results in bibliometrics is critical due to the field's relevance for research evaluation and research policy. We demonstrated how multiverse analysis can contribute to improving the transparency and credibility of bibliometric research. Whenever there is substantial model uncertainty (180 equally plausible models in our case), conventional robustness checks may be insufficient for the assessment of model robustness because they only cover a limited number of alternative model specifications. Multiverse analysis overcomes the limitations of conventional robustness checks by identifying and subsequently eliminating sources of model uncertainty.

In this study, we argued that credible research requires a thorough appraisal of modeling assumptions. However, there are modeling assumptions, which we could not investigate within the scope of this article. Specifically, we did not question the validity of the disruption index. Our conclusion regarding the team effect on disruptive research rests on the assumption that (some variants of) the disruption index are valid measures of disruptive research. The current state of research on the validity of disruption scores is mixed: Several studies support the validity of (specific variants of) the disruption index (Bittmann et al., 2022; Bornmann et al., 2020a; Bornmann & Tekles, 2021; Deng & Zeng, 2023; Wei et al., 2023), but the indicator also faces empirical and conceptual criticism (Bentley et al., 2023; Leibel et al., 2024; Petersen et al., 2024; Wang et al., 2023; Wu & Wu, 2019). Whether the disruption index is valid or not is unrelated to the research goal of this study, which is to illustrate how multiverse analysis may contribute to more robust bibliometric research.

Depending on the statistics software, the application of multiverse analysis may be limited to specific models. In this study, we mimicked the model specification used by Wu et al. (2019). As a result, our analyses share the same limitation: Fixed effects regression assume nested data, but this assumption is violated if publications appear in the publication records of multiple authors. As a result of this violation, the standard errors are underestimated in the fixed effects regressions. Models that could provide improved standard errors are currently not compatible with the *multivrs* module for Stata. Fortunately, our sample is large enough that even a multiplicative increase in the standard errors would have little effect on our research outcomes. Additionally, standard errors are of secondary importance for our research goals. Quantifying and reducing model variance require accurate coefficients, but not necessarily accurate standard errors.

We chose strict sample selection criteria to maximize metadata quality in testing computational model robustness. The fact that our sample is restricted to mostly natural sciences and medicine papers from 2000 to 2005 limits the generalizability of our findings. We acknowledge this limitation of our study while pointing out that generalizability and robustness are distinct qualities of statistical analyses. Model robustness addresses the variance of estimates across a set of defensible modelling assumptions. Generalizability concerns the question whether a pattern is universal or varies across subgroups of the target population. For example, the effect of team size on disruption scores may vary across time periods and disciplines. Testing whether the team effect describes a universal pattern in the production of scientific knowledge would require a replication of Wu et al. (2019) and Petersen et al. (2025), which is beyond the scope of this paper (and beyond the coverage offered by our bibliometric inhouse database).

Our analyses share a limitation with the studies of Wu et al. (2019) and Petersen et al. (2025): With observational data, it is not possible to control for the research conditions under which small teams and large teams conduct their scientific work. We note that the average size of research teams varies across countries and research institutions. For example, large teams are more prevalent in China than in other countries (Liu et al., 2021). It may be the case that small teams produce more disruptive research than large teams, not because they are small, but because they profit from national or institutional policies that support risky and creative research. Neither this study nor the studies of Wu et al. (2019) and Petersen et al. (2025) eliminate the possibility that the team effect is confounded by research conditions. This limitation illustrates the challenges that arise when estimating causal effects with observational data. Even though this study finds evidence of a robust team effect, the result needs to be interpreted with caution due to the limitations of non-experimental research designs.

# 7 Conclusion

Using multiverse analysis as a statistical tool for the investigation of influential modeling assumptions, we managed to reconcile the contradictory findings of Wu et al. (2019) and Petersen et al. (2025) regarding the question whether small teams have a critical role in producing disruptive research. On the one hand, we found support for the claim of Petersen et al. (2025) that controlling for citation inflation reduces the size of the team effect. On the other hand, we found robust evidence supporting the hypothesis of Wu et al. (2019) that small teams produce more disruptive research output than large teams. A multiverse analysis of plausible models showed that the size of the team effect varies substantially across model specifications: The difference in disruptive potential between small and large teams is either considerable or very small depending on the modeling assumptions.

Beyond the mechanisms through which research teams generate their research output, our results have methodological implications. Multiverse analysis provides computational measures of model robustness, but these are only fruitful in combination with well justified modeling assumptions. Our multiverse analysis demonstrates that the empirical results of studies based on the disruption index may depend significantly on the exact definition of the index. In other words, the calculation of disruption indices is a major source of model uncertainty, primarily because the modeling assumption that all citations indicate knowledge flow is violated in most citation networks. As noisy citation data is almost omnipresent in empirical bibliometric research, other bibliometric indices, e.g. interdisciplinarity measures (Wang & Schneider, 2020), may also generate substantial model uncertainty.

In this study, we demonstrated the dependency of empirical results on model specifications. To enhance the robustness of empirical results, it is standard practice in bibliometrics to test the inferential sturdiness of research outcomes with several alternative model specifications (with respect to indicators, control variables, etc.). However, the efficiency of this conventional approach to assessing model robustness is limited when there is large amount of model uncertainty. Investigating all plausible model specifications is only possible with multiverse analyses. As a first example of the application of multiverse analysis in bibliometrics, we investigated the relationship of team size and disruptive research output. Using multiverse analysis, we were able to explain and reconcile the divergent findings in previous research (Petersen et al., 2025; Wu et al., 2019). Future research could contribute to robust bibliometrics by using multiverse analysis to shed light on the hidden multiverse of bibliometric analyses (Leibel & Bornmann, 2024a).

# Appendix

Table 8: Complete regression table for the W-specification

| | Coefficient | SE | CI | | p |
|---|---|---|---|---|---|
| | | | LL | UL | |
| Team size | | | | | |
| 1 | Reference | | | | |
| 2 | -10.83 | 0.72 | -12.23 | -9.42 | < 0.001 |
| 3 | -13.35 | 0.70 | -14.72 | -11.98 | < 0.001 |
| 4 | -14.83 | 0.69 | -16.19 | -13.48 | < 0.001 |
| 5 | -16.34 | 0.69 | -17.69 | -14.99 | < 0.001 |
| 6 | -16.92 | 0.69 | -18.27 | -15.57 | < 0.001 |
| 7 | -18.39 | 0.69 | -19.74 | -17.03 | < 0.001 |
| 8 | -19.24 | 0.69 | -20.60 | -17.88 | < 0.001 |
| 9 | -20.08 | 0.70 | -21.45 | -18.71 | < 0.001 |
| 10 | -20.88 | 0.70 | -22.26 | -19.50 | < 0.001 |
| 11 | -22.02 | 0.72 | -23.43 | -20.62 | < 0.001 |
| 12 | -22.27 | 0.72 | -23.69 | -20.86 | < 0.001 |
| 13 | -23.39 | 0.75 | -24.86 | -21.93 | < 0.001 |
| 14 | -24.65 | 0.78 | -26.17 | -23.13 | < 0.001 |
| 15 | -23.66 | 0.81 | -25.25 | -22.08 | < 0.001 |
| Research field | | | | | |
| Biology | Reference | | | | |
| Chemistry | 0.41 | 0.48 | -0.52 | 1.34 | 0.389 |
| Engineering | 1.93 | 0.67 | 0.61 | 3.25 | 0.004 |
| Medicine | 0.30 | 0.14 | 0.04 | 0.57 | 0.025 |
| Multidisciplinary sciences | -5.85 | 0.28 | -6.41 | -5.30 | < 0.001 |
| Physical sciences | -0.01 | 0.57 | -1.12 | 1.10 | 0.985 |
| Other | 4.04 | 1.00 | 2.08 | 6.00 | < 0.001 |
| Publication year | | | | | |
| 2000 | Reference | | | | |
| 2001 | -0.91 | 0.20 | -1.30 | -0.52 | < 0.001 |
| 2002 | -0.94 | 0.19 | -1.32 | -0.57 | < 0.001 |
| 2003 | -0.23 | 0.19 | -0.60 | 0.14 | 0.217 |
| 2004 | -0.23 | 0.19 | -0.60 | 0.14 | 0.226 |
| 2005 | 0.03 | 0.19 | -0.34 | 0.40 | 0.879 |
| Constant | 67.15 | 0.70 | 65.79 | 68.52 | < 0.001 |
| $R^2$ | 0.01 | | | | |
| n | 469,590 | | | | |

*Note*. SE = standard error, CI = 95% confidence interval, LL = lower limit, UL = upper limit. Standard errors are clustered using author IDs.

Table 9: Complete regression table for the P-specification

| | Coefficient | SE | CI | | p |
|---|---|---|---|---|---|
| | | | LL | UL | |
| Team size | | | | | |
| 1 | Reference | | | | |
| 2 | -5.50 | 0.64 | -6.76 | -4.24 | < 0.001 |
| 3 | -6.04 | 0.63 | -7.26 | -4.81 | < 0.001 |
| 4 | -5.87 | 0.62 | -7.09 | -4.66 | < 0.001 |
| 5 | -5.51 | 0.62 | -6.72 | -4.29 | < 0.001 |
| 6 | -4.69 | 0.62 | -5.91 | -3.47 | < 0.001 |
| 7 | -4.95 | 0.63 | -6.19 | -3.71 | < 0.001 |
| 8 | -4.55 | 0.65 | -5.82 | -3.28 | < 0.001 |
| 9 | -4.10 | 0.67 | -5.42 | -2.78 | < 0.001 |
| 10 | -3.58 | 0.70 | -4.96 | -2.20 | < 0.001 |
| 11 | -4.28 | 0.77 | -5.78 | -2.78 | < 0.001 |
| 12 | -3.29 | 0.84 | -4.94 | -1.65 | < 0.001 |
| 13 | -4.26 | 0.97 | -6.15 | -2.37 | < 0.001 |
| 14 | -4.46 | 1.13 | -6.68 | -2.23 | < 0.001 |
| 15 | -3.17 | 1.27 | -5.66 | -0.68 | 0.013 |
| Research field | | | | | |
| Biology | Reference | | | | |
| Chemistry | 1.41 | 0.30 | 0.81 | 2.00 | < 0.001 |
| Engineering | -1.92 | 0.67 | -3.22 | -0.61 | 0.004 |
| Medicine | -0.73 | 0.17 | -1.07 | -0.38 | < 0.001 |
| Multidisciplinary sciences | 1.54 | 0.47 | 0.63 | 2.46 | 0.001 |
| Physical sciences | -1.05 | 0.31 | -1.67 | -0.44 | 0.001 |
| Other | -6.77 | 1.02 | -8.78 | -4.77 | < 0.001 |
| Publication year | | | | | |
| 2000 | Reference | | | | |
| 2001 | -0.40 | 0.32 | -1.02 | 0.22 | 0.205 |
| 2002 | 0.15 | 0.30 | -0.45 | 0.74 | 0.631 |
| 2003 | 0.82 | 0.29 | 0.25 | 1.40 | 0.005 |
| 2004 | 1.64 | 0.29 | 1.08 | 2.21 | < 0.001 |
| 2005 | 1.88 | 0.28 | 1.33 | 2.44 | < 0.001 |
| Reference counts (logarithmic) | 12.12 | 1.77 | 8.65 | 15.59 | < 0.001 |
| Squared reference counts (logarithmic) | 0.17 | 0.27 | -0.36 | 0.71 | 0.530 |
| Citation counts (logarithmic) | -11.25 | 0.23 | -11.70 | -10.81 | < 0.001 |
| Squared citation counts (logarithmic) | -0.69 | 0.04 | -0.78 | -0.61 | < 0.001 |
| Constant | 49.56 | 2.88 | 43.93 | 55.20 | < 0.001 |

Table 9: Complete regression table for the P-specification

| | |
|---|---|
| $R^2$ | 0.30 |
| n | 108,305 |

*Note*. SE = standard error, CI = 95% confidence interval, LL = lower limit, UL = upper limit.

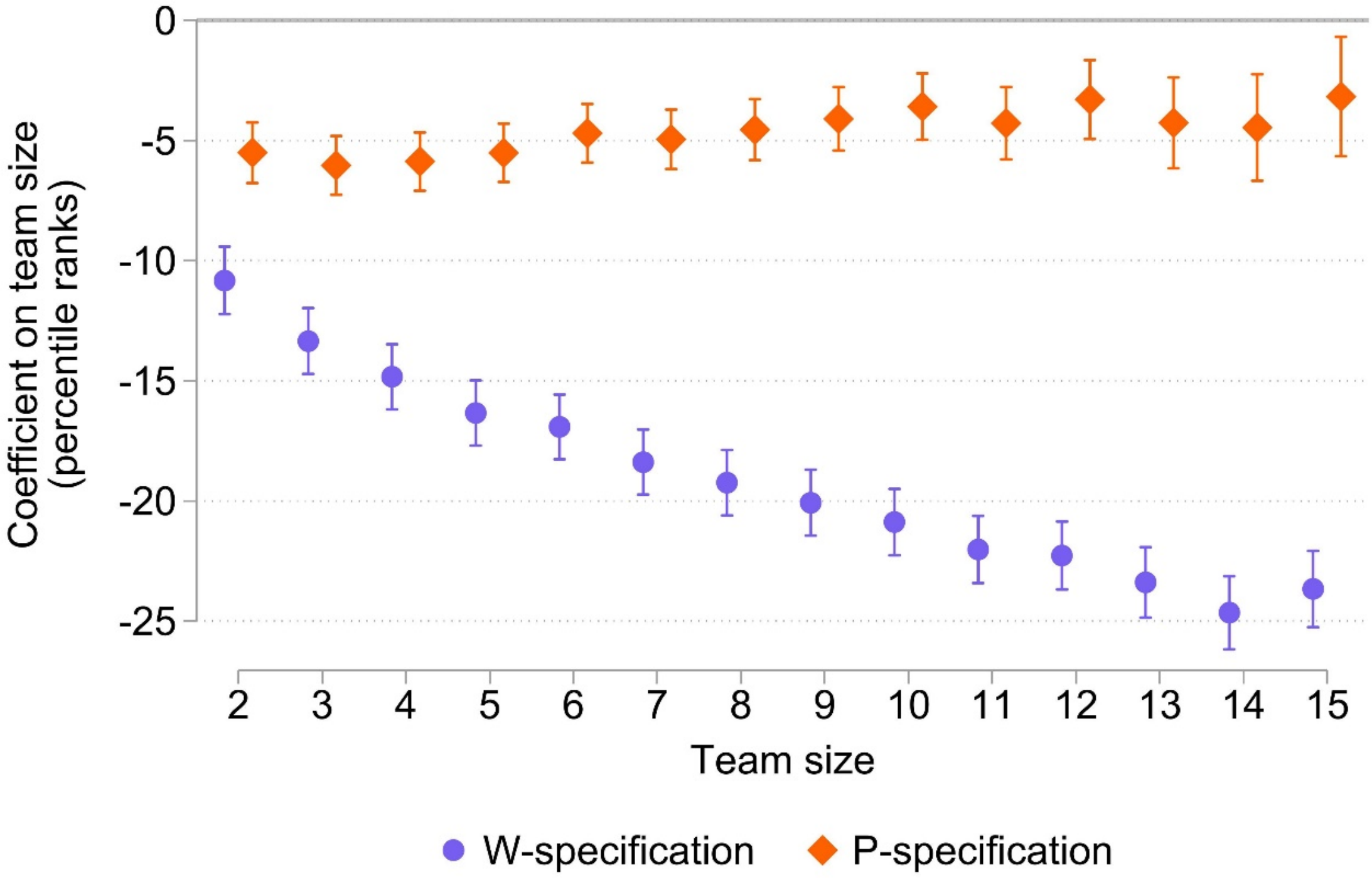

Figure 2: Coefficients from the model specifications proposed by Wu et al. (2019) and Petersen et al. (2025). The plot shows the coefficients of disruption percentile ranks on team size and the 95% confidence intervals. All coefficients are negative because the reference category (i.e. articles by only one author) have the highest average disruptions scores in the sample.

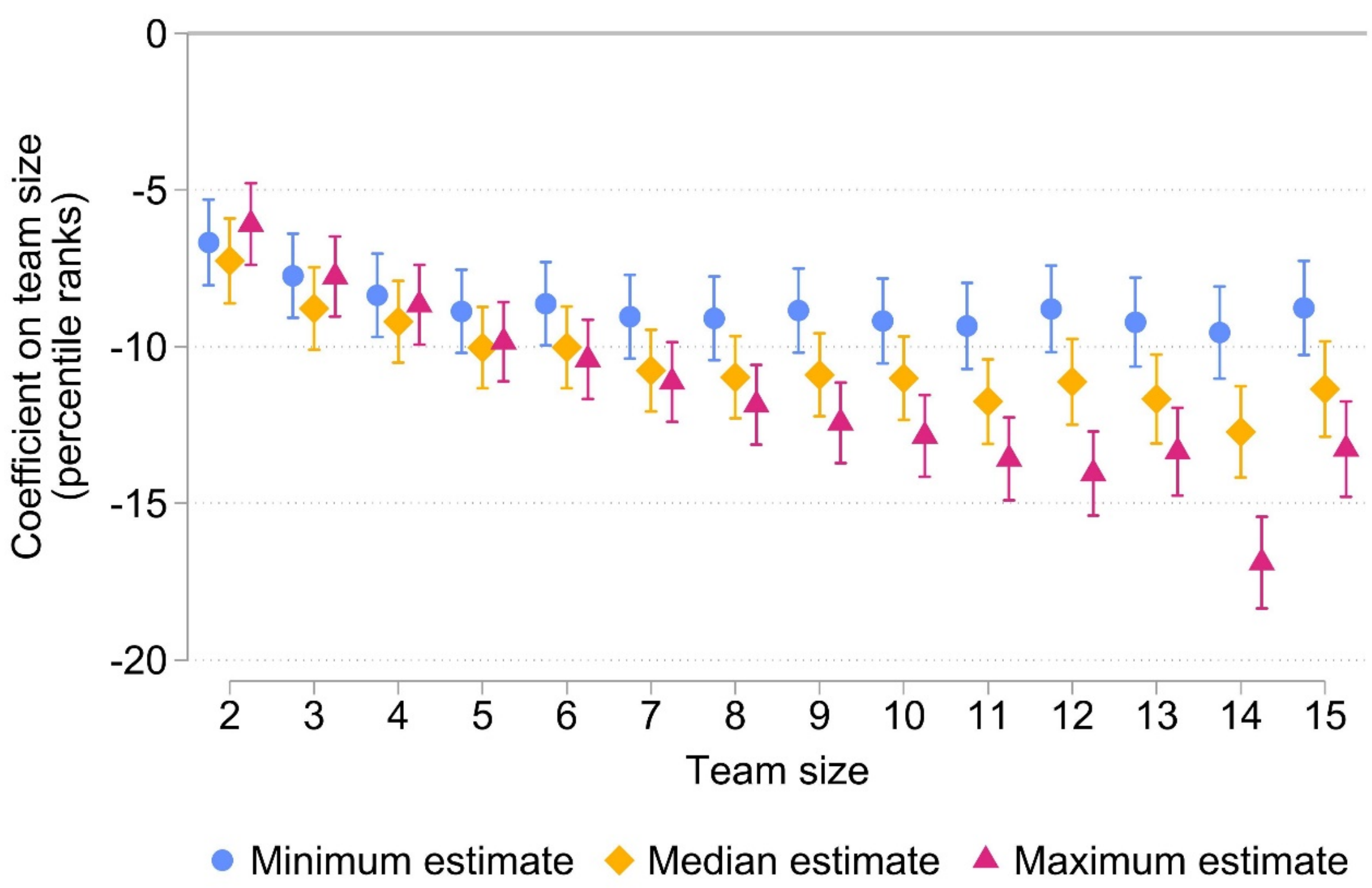

Figure 3: Coefficients from the model specifications that produce the minimum, median, and maximum estimates in the multiverse of 180 equally plausible models. The plot shows the coefficients of disruption percentile ranks on team size and the 95% confidence intervals. All coefficients are negative because the reference category (i.e. articles by only one author) have the highest average disruptions scores in the sample.

Table 10: Model space covering all possible combinations of model ingredients from the discussion between Wu et al. (2019) and Petersen et al. (2025)

| Dimension | | Specifications | |
|---|---|---|---|
| Operationalization | Disruption index | 1 – $DI_1$<br>2 – $DI_2$<br>3 – $DI_3$<br>4 – $DI_4$<br>5 – $DI_5$ | 5 |
| | Citation window | 1 – 5 years<br>2 – 10 years<br>3 – 15 years<br>4 – All citations up to 2023 | 4 |
| Covariates | Publication year | 1 – included<br>2 – excluded | 2 |
| | Discipline | 1 – included<br>2 – excluded | 2 |
| | Citation count | 1 – included<br>2 – excluded | 2 |
| | Reference count | 1 – included<br>2 – excluded | 2 |
| Exclusion criteria | Outliers | 1 – included<br>2 – excluded | 2 |
| | Minimum reference count | 1 – 1 cited reference<br>2 – 5 cited references<br>3 – 10 cited references | 3 |
| Model | Type of regression model | 1 – Regression<br>2 – Author fixed effects regression | 2 |
| | | | 3,840 |

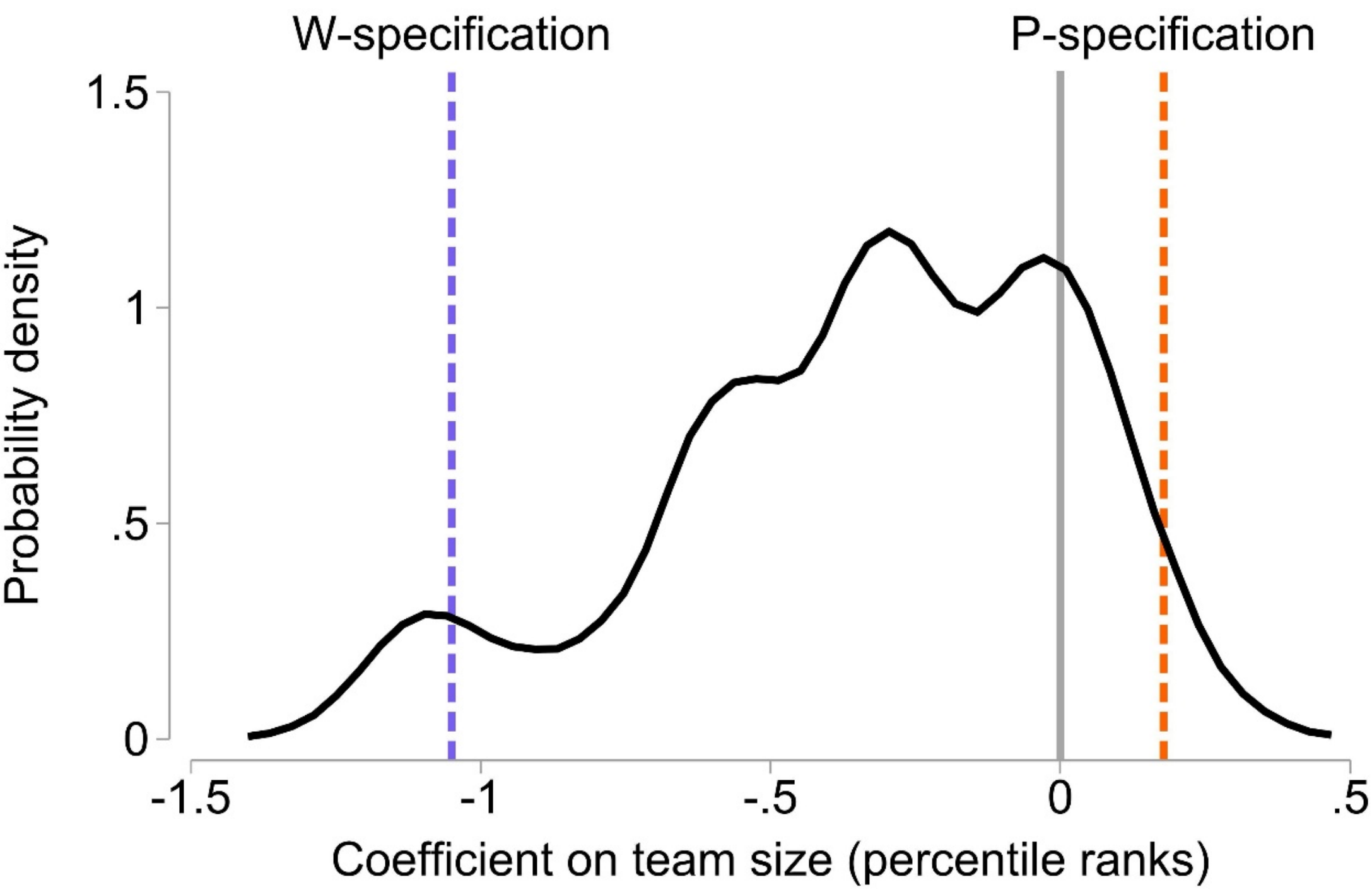

Figure 4: Probability density of estimates from 3,840 regression models. The coefficients obtained from the W-specification and the P-specification are highlighted. Effect sizes are measured in percentile ranks.

## Acknowledgements

We thank the authors of Wu et al. (2019) for providing their classification system of the Web of Science subject categories upon request.

## Competing interests

The authors have no competing interests.

## Funding information

This study uses ivsdb data for the bibliometric analyses. The ivsdb data are developed in cooperation between the Administrative Headquarters of the Max Planck Society (MPG) and the Information Retrieval Service for the institutes of the Chemistry, Physics and Technology (IVS-CPT) Section of the MPG. The data stem from a bibliometrics database provided by the German Kompetenznetzwerk Bibliometrie (KB, Competence Network Bibliometrics, funded since December 2008 by the Federal Ministry of Research, Technology and Space of Germany (BMFTR), currently under the grant numbers 16WIK2101A, 16WIK2101B, and 16WIK2101C, see: http://www.bibliometrie.info/). The data are derived from the Science Citation Index – Expanded (SCI-E), the Social Sciences Citation Index (SSCI), the Conference Proceedings Citation Index – Science (CPCI-S), the Conference Proceedings Citation Index – Social Science & Humanities (CPCI-SSH), and the Arts and Humanities Citation Index (AHCI), provided by Clarivate and updated in calendar week 17 of 2024.

## Data availability

We are unable to make the primary data available due to licensing restrictions. The results from the multiverse analysis (generated by the *multivrs* module) are available from https://doi.org/10.5281/zenodo.17790418

## Code availability

Code for the replication of Table 5, Table 6, Table 7, Figure 1, and Figure 4 is available from

https://doi.org/10.5281/zenodo.17790418

# References

Aksnes, D. W., & Sivertsen, G. (2019). A criteria-based assessment of the coverage of Scopus and Web of Science. *Journal of Data and Information Science*, *4*(1), 1-21. https://doi.org/10.2478/jdis-2019-0001

Angrist, J. D., & Pischke, J.-S. (2009). *Mostly harmless econometrics*. Princeton University Press. https://doi.org/10.1515/9781400829828

Auspurg, K. (2025). Robustness is better assessed with a few thoughtful models than with billions of regressions. *Proceedings of the National Academy of Sciences of the United States of America*, *122*(43), Article e2521917122. https://doi.org/10.1073/pnas.2521917122

Baas, J., Schotten, M., Plume, A., Côté, G., & Karimi, R. (2020). Scopus as a curated, high-quality bibliometric data source for academic research in quantitative science studies. *Quantitative Science Studies*, *1*(1), 377-386. https://doi.org/10.1162/qss_a_00019

Bentley, R. A., Valverde, S., Borycz, J., Vidiella, B., Horne, B. D., Duran-Nebreda, S., & O'Brien, M. J. (2023). Is disruption decreasing, or is it accelerating? *Advances in Complex Systems*, *26*(2), Article 2350006. https://doi.org/10.1142/s0219525923500066

Bischof, D. (2017). New graphic schemes for Stata: plotplain and plottig. *Stata Journal*, *17*(3), 748-759. https://doi.org/10.1177/1536867x1701700313

Bittmann, F., Tekles, A., & Bornmann, L. (2022). Applied usage and performance of statistical matching in bibliometrics: The comparison of milestone and regular papers with multiple measurements of disruptiveness as an empirical example. *Quantitative Science Studies*, *2*(4), 1246-1270. https://doi.org/10.1162/qss_a_00158

Bornmann, L., Devarakonda, S., Tekles, A., & Chacko, G. (2020a). Are disruption index indicators convergently valid? The comparison of several indicator variants with assessments by peers. *Quantitative Science Studies*, *1*(3), 1242-1259. https://doi.org/10.1162/qss_a_00068

Bornmann, L., Devarakonda, S., Tekles, A., & Chacko, G. (2020b). Disruptive papers published in *Scientometrics*: Meaningful results by using an improved variant of the disruption index originally proposed by Wu, Wang, and Evans (2019). *Scientometrics*, *123*(2), 1149-1155. https://doi.org/10.1007/s11192-020-03406-8

Bornmann, L., Ettl, C., & Leibel, C. (2024). In search of innovative potential: The challenge of measuring disruptiveness of research using bibliometric data. *Embo Reports*, *25*(7), 2837-2841. https://doi.org/10.1038/s44319-024-00177-8

Bornmann, L., Haunschild, R., & Mutz, R. (2021). Growth rates of modern science: A latent piecewise growth curve approach to model publication numbers from established and new literature databases. *Humanities & Social Sciences Communications*, *8*(1), Article 224. https://doi.org/10.1057/s41599-021-00903-w

Bornmann, L., & Tekles, A. (2019a). Disruption index depends on length of citation window. *Profesional De La Informacion*, *28*(2), Article e280207. https://doi.org/10.3145/epi.2019.mar.07

Bornmann, L., & Tekles, A. (2019b). Disruptive papers published in *Scientometrics*. *Scientometrics*, *120*(1), 331-336. https://doi.org/10.1007/s11192-019-03113-z

Bornmann, L., & Tekles, A. (2021). Convergent validity of several indicators measuring disruptiveness with milestone assignments to physics papers by experts. *Journal of Informetrics*, *15*(3), Article 101159. https://doi.org/10.1016/j.joi.2021.101159

Brüderl, J., & Ludwig, V. (2015). Fixed-effects panel regression. In H. Best & C. Wolf (Eds.), *The SAGE Handbook of Regression Analysis and Causal Inference* (pp. 327-358). SAGE. https://doi.org/10.4135/9781446288146

Cantone, G. G., & Tomaselli, V. (2024). Characterisation and calibration of multiversal methods. *Advances in Data Analysis and Classification*. https://doi.org/10.1007/s11634-024-00610-9

Chatfield, C. (1995). Model uncertainty, data mining and statistical inference. *Journal of the Royal Statistical Society. Series A (Statistics in Society)*, *158*(3), 419-466. https://doi.org/10.2307/2983440

Dai, C., Chen, Q., Wan, T., Liu, F., Gong, Y. B., & Wang, Q. F. (2021). Literary runaway: Increasingly more references cited per academic research article from 1980 to 2019. *Plos One*, *16*(8), Article e0255849. https://doi.org/10.1371/journal.pone.0255849

Davis, K. B., Mewes, M. O., Andrews, M. R., Vandruten, N. J., Durfee, D. S., Kurn, D. M., & Ketterle, W. (1995). Bose-Einstein condensation in a gas of sodium atoms. *Physical Review Letters*, *75*(22), 3969-3973. https://doi.org/10.1103/PhysRevLett.75.3969

Del Giudice, M., & Gangestad, S. W. (2021). A traveler's guide to the multiverse: Promises, pitfalls, and a framework for the evaluation of analytic decisions. *Advances in Methods and Practices in Psychological Science*, *4*(1), Article 2515245920954925. https://doi.org/10.1177/2515245920954925

Deng, N., & Zeng, A. (2023). Enhancing the robustness of the disruption metric against noise. *Scientometrics*, *128*(4), 2419–2428. https://doi.org/10.1007/s11192-023-04644-2

Elwert, F., & Winship, C. (2014). Endogenous selection bias: The problem of conditioning on a collider variable. *Annual Review of Sociology*, *40*, 31-53. https://doi.org/10.1146/annurev-soc-071913-043455

Funk, R. J., & Owen-Smith, J. (2017). A dynamic network measure of technological change. *Management Science*, *63*(3), 791-817. https://doi.org/10.1287/mnsc.2015.2366

Gelman, A., & Loken, E. (2014). The statistical crisis in science. *American Scientist*, *102*(6), 460-465. https://doi.org/10.1511/2014.111.460

Götz, M., Sarma, A., & O'Boyle, E. H. (2024). The multiverse of universes: A tutorial to plan, execute and interpret multiverses analyses using the R package *multiverse*. *International Journal of Psychology*, *59*(6), 1003-1014. https://doi.org/10.1002/ijop.13229

Hirsch, J. E. (2005). An index to quantify an individual's scientific research output. *Proceedings of the National Academy of Sciences of the United States of America*, *102*(46), 16569-16572. https://doi.org/10.1073/pnas.0507655102

Holst, V., Algaba, A., Tori, F., Wenmackers, S., & Ginis, V. (2024). *Dataset artefacts are the hidden drivers of the declining disruptiveness in science* [Preprint]. arXiv. https://doi.org/10.48550/arXiv.2402.14583

Huang, Y., Chen, L. X., & Zhang, L. (2020). Patent citation inflation: The phenomenon, its measurement, and relative indicators to temper its effects. *Journal of Informetrics*, *14*(2), Article 101015. https://doi.org/10.1016/j.joi.2020.101015

Jann, B. (2005). Making regression tables from stored estimates. *Stata Journal*, *5*(3), 288-308. https://doi.org/10.1177/1536867X0500500302

Jann, B. (2014). Plotting regression coefficients and other estimates. *Stata Journal*, *14*(4), 708-737. https://doi.org/10.1177/1536867x1401400402

Ke, Q., Ferrara, E., Radicchi, F., & Flammini, A. (2015). Defining and identifying Sleeping Beauties in science. *Proceedings of the National Academy of Sciences of the United States of America*, *112*(24), 7426-7431. https://doi.org/10.1073/pnas.1424329112

Leahey, E., Lee, J., & Funk, R. J. (2023). What types of novelty are most disruptive? *American Sociological Review*, *88*(3), 562–597. https://doi.org/10.1177/00031224231168074

Leamer, E. E. (1983). Let's take the con out of Econometrics. *American Economic Review*, *73*(1), 31-43. https://www.jstor.org/stable/1803924

Leamer, E. E. (1985). Sensitivity analyses would help. *American Economic Review*, *75*(3), 308-313. https://www.jstor.org/stable/1814801

Leibel, C., Aksnes, D. W., Tekles, A., & Bornmann, L. (2024). *Groundbreaking research and disruption: Empirical results on the correlation between assessments of groundbreaking research by peers and disruption index scores.* 28th International Conference on Science, Technology and Innovation Indicators, Berlin, Germany. https://doi.org/10.5281/zenodo.14170878

Leibel, C., & Bornmann, L. (2024a). Specification uncertainty: What the disruption index tells us about the (hidden) multiverse of bibliometric indicators. *Scientometrics*, *129*(12), 7971-7979. https://doi.org/10.1007/s11192-024-05201-1

Leibel, C., & Bornmann, L. (2024b). What do we know about the disruption index in scientometrics? An overview of the literature. *Scientometrics*, *129*(1), 601-639. https://doi.org/10.1007/s11192-023-04873-5

Leydesdorff, L., & Bornmann, L. (2021). Disruption indices and their calculation using web-of-science data: Indicators of historical developments or evolutionary dynamics? *Journal of Informetrics*, *15*(4), Article 101219. https://doi.org/10.1016/j.joi.2021.101219

Li, H. Y., Tessone, C. J., & Zeng, A. (2024). Productive scientists are associated with lower disruption in scientific publishing. *Proceedings of the National Academy of Sciences of the United States of America*, *121*(21), Article e2322462121. https://doi.org/10.1073/pnas.2322462121

Li, M. T., Livan, G., & Righi, S. (2024). Breaking down the relationship between disruption scores and citation counts. *Plos One*, *19*(12), Article e0313268. https://doi.org/10.1371/journal.pone.0313268

Liang, G., Lou, Y., & Hou, H. (2022). Revisiting the disruptive index: Evidence from the Nobel Prize-winning articles. *Scientometrics*, *127*(10), 5721-5730. https://doi.org/10.1007/s11192-022-04499-z

Lin, Y., Frey, C. B., & Wu, L. (2023). Remote collaboration fuses fewer breakthrough ideas. *Nature*, *623*(7989), 987-991. https://doi.org/10.1038/s41586-023-06767-1

Lin, Y., Li, L., & Wu, L. (2025a). *The disruption index measures displacement between a paper and its most cited reference* [Preprint]. arXiv. https://doi.org/10.48550/arXiv.2504.04677

Lin, Y., Li, L., & Wu, L. (2025b). Team size and its negative impact on the disruption index. *Journal of Informetrics*, *19*(3), Article 101678. https://doi.org/10.1016/j.joi.2025.101678

Lin, Z., Yin, Y., Liu, L., & Wang, D. (2023). SciSciNet: A large-scale open data lake for the science of science research. *Scientific Data*, *10*(1), Article 315. https://doi.org/10.1038/s41597-023-02198-9

Liu, L. L., Yu, J. F., Huang, J. M., Xia, F., & Jia, T. (2021). The dominance of big teams in China's scientific output. *Quantitative Science Studies*, *2*(1), 350-362. https://doi.org/10.1162/qss_a_00099

Lundberg, I., Johnson, R., & Stewart, B. M. (2021). What is your estimand? Defining the target quantity connects statistical evidence to theory. *American Sociological Review*, *86*(3), 532-565. https://doi.org/10.1177/00031224211004187

Merton, R. K. (1988). The Matthew effect in science, II: Cumulative advantage and the symbolism of intellectual property. *Isis*, *79*(4), 606-623. http://www.jstor.org/stable/234750

Muñoz, J., & Young, C. (2018). We ran 9 billion regressions: Eliminating false positives through computational model robustness. *Sociological Methodology*, *48*(1), 1-33. https://doi.org/10.1177/0081175018777988

Nicolaisen, J., & Frandsen, T. F. (2021). Number of references: A large-scale study of interval ratios. *Scientometrics*, *126*(1), 259-285. https://doi.org/10.1007/s11192-020-03764-3
Opthof, T., & Leydesdorff, L. (2010). *Caveats* for the journal and field normalizations in the CWTS ("Leiden") evaluations of research performance. *Journal of Informetrics*, *4*(3), 423-430. https://doi.org/10.1016/j.joi.2010.02.003
Pan, R. K., Petersen, A. M., Pammolli, F., & Fortunato, S. (2018). The memory of science: Inflation, myopia, and the knowledge network. *Journal of Informetrics*, *12*(3), 656-678. https://doi.org/10.1016/j.joi.2018.06.005
Park, M., Leahey, E., & Funk, R. J. (2023). Papers and patents are becoming less disruptive over time. *Nature*, *613*(7942), 138-144. https://doi.org/10.1038/s41586-022-05543-x
Park, M., Leahey, E., & Funk, R. J. (2025). *Robust evidence for declining disruptiveness: Assessing the role of zero-backward-citation works* [Preprint]. arXiv. https://doi.org/10.48550/arXiv.2503.00184
Patel, C. J., Burford, B., & Ioannidis, J. P. A. (2015). Assessment of vibration of effects due to model specification can demonstrate the instability of observational associations. *Journal of Clinical Epidemiology*, *68*(9), 1046-1058. https://doi.org/10.1016/j.jclinepi.2015.05.029
Petersen, A. M., Arroyave, F. J., & Pammolli, F. (2024). The disruption index is biased by citation inflation. *Quantitative Science Studies*, *5*(4), 936-953. https://doi.org/10.1162/qss_a_00333
Petersen, A. M., Arroyave, F. J., & Pammolli, F. (2025). The disruption index suffers from citation inflation: Re-analysis of temporal CD trend and relationship with team size reveal discrepancies. *Journal of Informetrics*, *19*(1), Article 101605. https://doi.org/10.1016/j.joi.2024.101605
Petersen, A. M., Pan, R. K., Pammolli, F., & Fortunato, S. (2019). Methods to account for citation inflation in research evaluation. *Research Policy*, *48*(7), 1855-1865. https://doi.org/10.1016/j.respol.2019.04.009
Ruan, X., Lyu, D., Gong, K., Cheng, Y., & Li, J. (2021). Rethinking the disruption index as a measure of scientific and technological advances. *Technological Forecasting and Social Change*, *172*, Article 121071. https://doi.org/10.1016/j.techfore.2021.121071
Sánchez-Gil, S., Gorraiz, J., & Melero-Fuentes, D. (2018). Reference density trends in the major disciplines. *Journal of Informetrics*, *12*(1), 42-58. https://doi.org/https://doi.org/10.1016/j.joi.2017.11.003
Schmidt, M., Rimmert, C., Stephen, D., Lenke, C., Donner, P., Gärtner, S., Taubert, N., Bausenwein, T., & Stahlschmidt, S. (2025). The data infrastructure of the German Kompetenznetzwerk Bibliometrie: An enabling intermediary between raw data and analysis. *Quantitative Science Studies*, *6*, 1129-1146. https://doi.org/10.1162/QSS.a.20
Sheng, L., Lyu, D., Ruan, X., Shen, H., & Cheng, Y. (2023). The association between prior knowledge and the disruption of an article. *Scientometrics*, *128*(8), 4731–4751. https://doi.org/10.1007/s11192-023-04751-0
Simonsohn, U., Simmons, J. P., & Nelson, L. D. (2020). Specification curve analysis. *Nature Human Behaviour*, *4*(11), 1208-1214. https://doi.org/10.1038/s41562-020-0912-z
Steegen, S., Tuerlinckx, F., Gelman, A., & Vanpaemel, W. (2016). Increasing transparency through a multiverse analysis. *Perspectives on Psychological Science*, *11*(5), 702-712. https://doi.org/10.1177/1745691616658637
Tahamtan, I., & Bornmann, L. (2018a). Core elements in the process of citing publications: Conceptual overview of the literature. *Journal of Informetrics*, *12*(1), 203-216. https://doi.org/10.1016/j.joi.2018.01.002

Tahamtan, I., & Bornmann, L. (2018b). Creativity in science and the link to cited references: Is the creative potential of papers reflected in their cited references? *Journal of Informetrics*, *12*(3), 906-930. https://doi.org/10.1016/j.joi.2018.07.005

Tahamtan, I., & Bornmann, L. (2019). What do citation counts measure? An updated review of studies on citations in scientific documents published between 2006 and 2018. *Scientometrics*, *121*(3), 1635-1684. https://doi.org/10.1007/s11192-019-03243-4

Wang, Q., & Schneider, J. W. (2020). Consistency and validity of interdisciplinarity measures. *Quantitative Science Studies*, *1*(1), 239-263. https://doi.org/10.1162/qss_a_00011

Wang, S., Ma, Y., Mao, J., Bai, Y., Liang, Z., & Li, G. (2023). Quantifying scientific breakthroughs by a novel disruption indicator based on knowledge entities. *Journal of the Association for Information Science and Technology*, *74*(2), 150-167. https://doi.org/10.1002/asi.24719

Wei, C., Li, J., & Shi, D. (2023). Quantifying revolutionary discoveries: Evidence from Nobel prize-winning papers. *Information Processing & Management*, *60*(3), Article 103252. https://doi.org/10.1016/j.ipm.2022.103252

Western, B. (1996). Vague theory and model uncertainty in macrosociology. *Sociological Methodology*, *26*, 165-192. https://doi.org/10.2307/271022

Wu, L., Wang, D., & Evans, J. A. (2019). Large teams develop and small teams disrupt science and technology. *Nature*, *566*(7744), 378-382. https://doi.org/10.1038/s41586-019-0941-9

Wu, S., & Wu, Q. (2019). *A confusing definition of disruption* [Preprint]. SocArXiv. https://doi.org/10.31235/osf.io/d3wpk

Wuchty, S., Jones, B. F., & Uzzi, B. (2007). The increasing dominance of teams in production of knowledge. *Science*, *316*(5827), 1036-1039. https://doi.org/doi:10.1126/science.1136099

Young, C. (2009). Model uncertainty in sociological research: An application to religion and economic growth. *American Sociological Review*, *74*(3), 380-397. https://doi.org/10.1177/000312240907400303

Young, C. (2018). Model uncertainty and the crisis in science. *Socius*, *4*, 1-7. https://doi.org/10.1177/2378023117737206

Young, C., & Holsteen, K. (2017). Model uncertainty and robustness: A computational framework for multimodel analysis. *Sociological Methods & Research*, *46*(1), 3-40. https://doi.org/10.1177/0049124115610347